\documentclass[a4paper,onecolumn,11pt,accepted=2024-01-15]{quantumarticle}
\pdfoutput=1
\usepackage{geometry}
\usepackage[utf8]{inputenc}
\usepackage[english]{babel}
\usepackage[T1]{fontenc}
\usepackage{graphicx,graphics,times,bm,bbm,bbold,amssymb,soul}
\usepackage{amsthm,amsmath,amsfonts,dsfont,color,xcolor,dcolumn}
\usepackage{hyperref}
\usepackage{cleveref}
\usepackage{tikz}
\usepackage{lipsum}

\makeatletter
\newcommand{\xdashrightarrow}[2][]{\ext@arrow
0359\rightarrowfill@@{#1}{#2}}
\def\rightarrowfill@@{\arrowfill@@\relax\relbar\rightarrow}
\def\arrowfill@@#1#2#3#4{%
  $\m@th\thickmuskip0mu\medmuskip\thickmuskip\thinmuskip\thickmuskip
   \relax#4#1
   \xleaders\hbox{$#4#2$}\hfill
   #3$%
}
\makeatother

\usepackage{tabularx,ragged2e}

\DeclareMathOperator{\rk}{rk}

\DeclareMathOperator{\brk}{brk}

\DeclareMathOperator{\W}{W}
\DeclareMathOperator{\GHZ}{GHZ}
\DeclareMathOperator{\LL}{L}
\DeclareMathOperator{\M}{M}
\DeclareMathOperator{\N}{N}
\DeclareMathOperator{\Y}{Y}

\DeclareMathOperator{\Span}{Span}

\newtheorem{theorem}{Theorem}
\newtheorem{remark}{Remark}
\newtheorem{lemma}{Lemma}
\newtheorem{proposition}{Proposition}
\newtheorem{corollary}{Corollary}
\newtheorem{conjecture}{Conjecture}
\newtheorem{definition}{Definition}

\begin{document}

\title{Persistent Tensors and Multiqudit Entanglement \mbox{Transformation}}

\author{Masoud Gharahi}
\email[]{masoud.gharahi@gmail.com}
\email[]{masoud.gharahighahi@unicam.it}
\affiliation{QSTAR, INO-CNR and LENS, Largo Enrico Fermi 2, 50125 Firenze, Italy}
\orcid{0000-0002-7515-6179}
\homepage{https://sites.google.com/view/masoudgharahi}

\author{Vladimir Lysikov}
\email[]{vladimir.lysikov@rub.de}
\affiliation{Ruhr University Bochum, 44801 Bochum, Germany}
\orcid{0000-0002-7816-6524}


\begin{abstract}
We construct a lower bound of the tensor rank for a new class of tensors, which we call \emph{\mbox{persistent} tensors}. We present three specific families of persistent tensors, of which the lower bound is tight. We show that there is a chain of degenerations between these three families of minimal-rank persistent tensors that can be used to study the entanglement transformation between them. In addition, we show that these three families of persistent tensors are indeed different generalizations of multiqubit $\W$ states within multiqudit systems and are geometrically in the orbit closure of multiqudit $\GHZ$ states. Consequently, we show that one can obtain every one of the generalizations of $\W$ state from a multiqudit $\GHZ$ state via asymptotic Stochastic Local Operations and Classical Communication (SLOCC) with rate one. Finally, we extend the obtained lower bound of the tensor rank to direct sums with persistent summands and to even more general combinations of tensors, which we call \emph{block pyramidal tensors}. As a result, we show that the tensor rank is multiplicative under the Kronecker and tensor products of minimal-rank persistent tensors with the $\GHZ$ tensor.
\end{abstract}
\maketitle

\section{Introduction}\label{sec.i}
Entanglement, as one of the pillars of quantum information theory, is a fundamental resource for tasks that cannot be performed with the use of classical resources \cite{HHHH09}. Therefore, understanding and characterizing the entanglement in multipartite quantum states is one of the major goals. Actually, this was done for bipartite systems both in the Local Operation and Classical Communication (LOCC) and Stochastic LOCC (SLOCC) paradigms by developing entanglement monotones. However, much less is known about three or more parties. For example, while there is a complete entanglement classification of three-qubit systems under SLOCC equivalence \cite{DVC00,ABLS01}, we know a partial classification of three-qudit systems \cite{Nurmiev1,Nurmiev2,BLTV04,HJ16,GM21}.

One of the fundamental problems in quantum information theory concerns the interconversion between different resources by the SLOCC and asymptotic SLOCC. Using the Schmidt rank, one can characterize the SLOCC convertibility of bipartite systems. In fact, a bipartite quantum state is SLOCC convertible to another bipartite quantum state iff the Schmidt rank of the initial state is not smaller than that of the latter (notation: $|\psi\rangle\xrightarrow[]{\text{SLOCC}}|\varphi\rangle\Leftrightarrow\rk_S(\psi)\geq\rk_S(\varphi)$). A generalization of Schmidt rank in multipartite systems is the tensor rank.
Another tool relevant to the SLOCC entanglement transformation is the tensor border rank (border rank, for short). The border rank of a tensor $\mathcal{T}$ is defined as the smallest $r$ such that $\mathcal{T}$ is a limit of tensors of rank $r$. Both the tensor rank and the border rank have been extensively studied in algebraic complexity theory \cite{BCS97} and algebraic geometry \cite{Landsberg}. Recently, connections have been discovered between algebraic complexity theory, algebraic geometry, asymptotic SLOCC transformations, and SLOCC equivalence. \cite{CDS08,YCGD10,CDS10,CCDJW10,YGD14,VC15,VC17,GMO20}. An important property of the tensor rank is that it is an SLOCC monotone, that is, if a source quantum state $|\psi\rangle$ can be transformed into a target quantum state $|\varphi\rangle$ via SLOCC, then the tensor rank of the source is not smaller than that of the target (notation: $|\psi\rangle\xrightarrow[]{\text{SLOCC}}|\varphi\rangle\Rightarrow\rk(\psi)\geq\rk(\varphi)$). Although in general the inverse is not necessarily true, as in Ref. \cite{CDS08} it has been shown that a $\GHZ$-equivalent state (a state in the $\GHZ$ orbit) $|\psi_{\GHZ}\rangle$ can be transformed into another state $|\varphi\rangle$ iff the tensor rank of the $\GHZ$-equivalent state is not smaller than that of the latter, i.e., $|\psi_{\GHZ}\rangle\xrightarrow[]{\text{SLOCC}}|\varphi\rangle\Leftrightarrow\rk(\psi_{\GHZ})\geq\rk(\varphi)$. On the other hand, it is well known that a $\GHZ$ state cannot be transformed into a $\W$ state by SLOCC \cite{DVC00}, as they belong to distinct entanglement classes of multiqubit states, but one can asymptotically produce a $\W$-equivalent state from a $\GHZ$-equivalent state with rate arbitrarily close to one (see Refs. \cite{GMO20,VC15,GM21} for theory and Ref. \cite{WRZ05} for an experimental way). Actually, the reason is that the tensor rank of the multiqubit $\GHZ$ state is less than the tensor rank of the multiqubit $\W$ state, but the border rank of both of them is the same (geometrically, the $\W$ state is in the orbit closure of the $\GHZ$ state; see Ref. \cite{GMO20}). This phenomenon is known as degeneration in algebraic complexity theory \cite{BCS97} and algebraic geometry \cite{Landsberg} and is related to the asymptotic SLOCC transformation in entanglement theory \cite{GMO20,VC15,VC17,GM21}. Indeed, the border rank also has the same property as the tensor rank that is SLOCC monotone, i.e., a target quantum state $|\varphi\rangle$ can be approximated from a source quantum state $|\psi\rangle$ via SLOCC, then the border rank of the source is not smaller than that of the target (notation: $|\psi\rangle\xdashrightarrow[]{\text{SLOCC}}|\varphi\rangle\Rightarrow\brk(\psi)\geq\brk(\varphi)$). Interestingly, if a target quantum state $|\varphi\rangle$ can be obtained approximated from a $\GHZ$-equivalent state $|\psi_{\GHZ}\rangle$, then the border rank of the $\GHZ$-equivalent state is not smaller than that of the target, i.e., $|\psi_{\GHZ}\rangle\xdashrightarrow[]{\text{SLOCC}}|\varphi\rangle\Leftrightarrow\brk(\psi_{\GHZ})\geq\brk(\varphi)$.

Unlike Schmidt rank in the bipartite case, tensor rank is NP-hard to compute~\cite{Haastad90}.
Exact value of tensor rank is often not known even for simple tensors.
In this work, we introduce a new class of tensors in $\otimes_{i=1}^n\mathbbm{C}^{d_i}$ that we call \emph{persistent tensors} and construct a lower bound for their tensor rank. Furthermore, we show that the obtained tensor rank lower bound can be extended to direct sums of persistent tensors and to even more general combinations of tensors, which we call \emph{block pyramidal tensors}.
As a consequence, we show that the tensor rank is multiplicative under the Kronecker and the tensor product of minimal-rank persistent tensors with the $\GHZ$ tensor.
This allows us to determine the rank of the  product of $\GHZ$ and $\W$ tensors, answering an open question posed in Ref.~\cite{CF18}.
Our results are can be applied to other problems with similar structure, for example, we obtain the exact value for the tensor rank of the Kronecker product and the tensor product of $\GHZ$ with two copies of $\W$.

We also present three specific families of tensors in this class, which we denote by $\mathcal{L}(d,n)$, $\mathcal{M}(d, n)$, and $\mathcal{N}(d,n)$ (their corresponding $n$-qudit states we call the $\LL$, $\M$, and $\N$ states). $\LL$, $\M$, and $\N$ states can be seen as different generalizations of the multiqubit $\W$ state within multiqudit systems. In fact, $\mathcal{W}_n=\mathcal{L}(2,n)=\mathcal{M}(2,n)=\mathcal{N}(2,n)$. Furthermore, the Kronecker square of $\W$ is unitarily equivalent to the $\M$ state, i.e., $\mathcal{W}_n\boxtimes\mathcal{W}_n=\mathcal{M}(4, n)$. We show that the tensor rank of $\mathcal{L}(d,n)$, $\mathcal{M}(d, n)$, and $\mathcal{N}(d,n)$ coincides with the lower bound for persistent tensors. Then, we give a chain of degenerations between $\mathcal{L}(d,n)$, $\mathcal{M}(d,n)$, and $\mathcal{N}(d,n)$ that can be used to study the entanglement transformation between them. In addition, we prove that the border rank of these three families of minimal-rank persistent tensors is equal to $d$. This will reveal the fact that geometrically multiqudit $\LL$, $\M$, and $\N$ states are in the orbit closure of multiqudit $\GHZ$ states. Consequently, we show that a multiqudit $\GHZ$-equivalent state can be transformed into each of the multiqudit $\LL$-, $\M$-, and $\N$-equivalent states by asymptotic SLOCC with rate one. 


\section{Preliminaries}\label{sec.ii}


\subsection{Multipartite quantum states as multipartite tensors}
A state of a multipartite quantum system can be considered as a multipartite tensor in the tensor product of Hilbert spaces of each individual subsystem. Let $\mathcal{H}_n^{\mathbf{d}}=\otimes_{i=1}^{n}\mathbbm{C}^{d_i}$ be the Hilbert space representing the state space of an $n$-partite quantum system where $\mathbf{d}=(d_1,\ldots,d_n)$ indicates the dimensions of the Hilbert spaces of each individual subsystem.

There are two different notions of product for tensors. Suppose that we have two tensors $\mathcal{T}_1 \in\mathcal{H}_{n_1}^{\mathbf{d}}=\otimes_{i=1}^{n_1}\mathbbm{C}^{d_i}$ and $\mathcal{T}_2\in \mathcal{H}_{n_2}^{\mathbf{d'}}=\otimes_{i=1}^{n_2}\mathbbm{C}^{d'_i}$ corresponding to two multipartite quantum systems, the first with $n_1$ parties and the second with $n_2$ parties, respectively. Assume $n_1\leq n_2$ (without any loss of generality). The first product is the tensor product that corresponds to an ($n_1+n_2$)-partite system, and we denote it by $\mathcal{T}_1\otimes\mathcal{T}_2\in\mathcal{H}_{n_1}^{\mathbf{d}}\otimes\mathcal{H}_{n_2}^{\mathbf{d'}}$. The second product is the Kronecker product, which corresponds to an $n_2$-partite system in which the first $n_1$ parties share both states.
We denote the Kronecker product by $\mathcal{T}_1\boxtimes\mathcal{T}_2 \in \mathcal{H}_{n_1}^{\mathbf{d}}\boxtimes\mathcal{H}_{n_2}^{\mathbf{d'}}$. Here $\mathcal{H}_{n_1}^{\mathbf{d}}\otimes\mathcal{H}_{n_2}^{\mathbf{d'}}=(\otimes_{i=1}^{n_1}\mathbbm{C}^{d_i})\otimes(\otimes_{j=1}^{n_2}\mathbbm{C} ^{d'_j})$ and $\mathcal{H}_{n_1}^{\mathbf{d}}\boxtimes\mathcal{H}_{n_2}^{\mathbf{d'}}=(\otimes_{i=1}^{n_1}\mathbbm{C}^{d_i d'_i})\otimes(\otimes_{j=n_1+1}^{n_2}\mathbbm{C}^{d'_j})$.

We also need the notion of direct sum of tensors.
Let $\mathcal{T}_1 \in\mathcal{H}_{n}^{\mathbf{d}}=\otimes_{i=1}^{n}\mathbbm{C}^{d_i}$ and $\mathcal{T}_2\in \mathcal{H}_{n}^{\mathbf{d'}}=\otimes_{i=1}^{n}\mathbbm{C}^{d'_i}$ be two tensors with the same number of factors.
By considering the spaces $\mathbbm{C}^{d_i}$ and $\mathbbm{C}^{d'_i}$ as two summands of $(\mathbbm{C}^{d_i} \oplus \mathbbm{C}^{d'_i}) \cong \mathbbm{C}^{d_i + d'_i}$,
we can embed $\mathcal{T}_1$ and $\mathcal{T}_2$ into a larger Hilbert space $\mathcal{H}_{n}^{\mathbf{d}+\mathbf{d'}} \cong \otimes_{i=1}^{n}(\mathbbm{C}^{d_i} \oplus \mathbbm{C}^{d'_i})$.
The direct sum $\mathcal{T}_1 \oplus \mathcal{T}_2 \in \mathcal{H}_{n}^{\mathbf{d}+\mathbf{d'}}$ is the sum of the two tensors embedded in this way.

In this paper, we take $\{|j\rangle\mid j\in\mathbbm{Z}_d\}$ as the computational basis of $\mathbbm{C}^d$. We do not distinguish multipartite quantum states from the tensors that represent them. We denote specific tensors by calligraphic capital letters. It should be noted that we do not consider the normalization of the quantum states, since all properties that we work with can be defined for tensors in general and are invariant under scaling.

The state of a composite system is always expressible as a superposition of tensor products of the states of individual subsystems. A quantum state is called fully separable (or unentangled) if it can be written as a tensor product of individual subsystem states, i.e., $|\psi\rangle=|\varphi_1\rangle\otimes\cdots\otimes|\varphi_n\rangle$. Therefore, it is desirable to characterize the entanglement in a composite system. The tensor rank is a good tool for this purpose. The rank of a tensor $\mathcal{T}$ is defined as the minimum number of simple tensors (fully separable states) that sum up to $\mathcal{T}$. The following is a concrete definition of tensor rank.
\begin{definition}\label{def:rank}
Let $\mathcal{T}\in V_1\otimes\cdots\otimes V_n$ be a tensor where each $V_i$ is a vector space. The tensor rank of $\mathcal{T}$ is defined as follows
\begin{equation}\label{rank}
\rk(\mathcal{T})=\min\Big\{r~\big|~\mathcal{T}=\sum_{p=1}^{r}v_1^{(p)}\otimes\cdots\otimes v_n^{(p)}\,,~\text{for some}~v_i^{(p)}\in V_i\Big\}\,.
\end{equation}
\end{definition}
If $n=2$, the tensor rank of $\mathcal{T}$ is equal to the matrix rank of the linear map $\mathcal{T}\colon V_1^{\vee}\to V_2$ ($V_1^{\vee}$ is the dual of $V_1$); in this sense, it extends the notion of the rank of a matrix in algebra \cite{Bourbaki}.

The border rank of a tensor $\mathcal{T}$ is defined as follows.
\begin{definition}\label{def:brank}
Let $\mathcal{T}\in V_1\otimes\cdots\otimes V_n$ be a tensor where each $V_i$ is a vector space. The border rank of $\mathcal{T}$ is the smallest $r$ such that $\mathcal{T}$ is a limit of tensors of rank $r$, i.e.,
\begin{equation}\label{brank}
\brk(\mathcal{T})=\min\Big\{r~\big|~\mathcal{T}=\lim_{\varepsilon\to 0}\mathcal{T}_{\varepsilon}\,,\text{s.t.}~\forall\varepsilon\neq0,~\rk(\mathcal{T}_{\varepsilon})=r\Big\}\,.
\end{equation}
\end{definition}
Clearly, $\brk(\mathcal{T})\leq\rk(\mathcal{T})$.

For example, consider the $n$-qubit $\W$ and $\GHZ$ states. An $n$-qubit $\W$ state, i.e.,
\begin{equation}\label{W}
\mathcal{W}_n=\sum_{\mathfrak{p}\in\mathfrak{S}_n}\mathfrak{p}\big\{|0\rangle^{\otimes(n-1)}|1\rangle\big\}=\sum_{i=0}^{n-1}|0\rangle^{\otimes(n-i-1)}|1\rangle|0\rangle^{\otimes{i}}\,,
\end{equation}
where $\mathfrak{p}$ denotes non-redundant elements of symmetric group $\mathfrak{S}_n$, corresponds to a symmetric tensor in $\otimes^{n}\mathbbm{C}^2$ and its tensor rank and border rank are known to be $\rk(\mathcal{W}_n)=n\,$\footnote{A complete proof seems to be missing in the literature. In Ref. \cite[Theorem 3]{CCDJW10} the tensor rank of multiqubit Dicke states is presented. The proof is based on induction and the base case is cited to Ref. \cite{DVC00}. While the techniques from~\cite{DVC00} can be used to infer that the tensor rank of $n$-qubit $\W$ state is $n$, it does not seem clear at first sight.}
and $\brk(\mathcal{W}_n)=2$, respectively. We will prove this fact that the tensor rank of a $n$-qubit $\W$ state is indeed $n$. A generalized $n$-qudit $\GHZ$ state, i.e.,
\begin{equation}\label{GHZ}
\mathcal{G}(d,n)=\sum_{j=0}^{d-1}|j\rangle^{\otimes{n}}\,,
\end{equation}
corresponds to a symmetric tensor in $\otimes^{n}\mathbbm{C}^d$ and its tensor rank and border rank are $\rk(\mathcal{G}(d,n))=\brk(\mathcal{G}(d,n))=d$.

It is known that a multiqudit $\GHZ$-equivalent state can be transformed into a quantum state $|\psi\rangle$ iff $d \geq \rk(|\psi\rangle)$ \cite{CDS08}. 
Therefore, the tensor rank of a quantum state can be characterized as follows
\begin{equation}\label{rank-SLOCC}
\rk(\mathcal{T})=\min\Big\{d~\big|~\mathcal{G}(d,n)\xrightarrow[]{\text{SLOCC}}\mathcal{T}\Big\}\,.
\end{equation}
Similarly, the border rank of a quantum state is the smallest $d$ such that a multiqudit $\GHZ$-equivalent state degenerates into it.
\begin{equation}\label{brank-degeneration}
\brk(\mathcal{T})=\min\Big\{d~\big|~\mathcal{G}(d,n)\xdashrightarrow[]{\text{SLOCC}}\mathcal{T}\Big\}\,.
\end{equation}

Now, assume that $|{\rm{S}}\rangle$ and $|{\rm{T}}\rangle$ are two quantum states in Hilbert spaces $\mathcal{H}$ and $\mathcal{H}'$ whose tensor ranks are $\rk(\mathcal{S})$ and $\rk(\mathcal{T})$, respectively. Then $\mathcal{T}\boxtimes\mathcal{S}\in\mathcal{H}\boxtimes\mathcal{H}'$, $\mathcal{T}\otimes\mathcal{S}\in\mathcal{H}\otimes\mathcal{H}'$, and we have the following inequalities
\begin{equation}\label{rank-inequalities}
\rk(\mathcal{S}\boxtimes\mathcal{T})\leq\rk(\mathcal{S}\otimes\mathcal{T})\leq\rk(\mathcal{S})\rk(\mathcal{T})\,.
\end{equation}
These operations (the Kronecker product and the tensor product) can be applied to the study of the (asymptotic) SLOCC interconversion between multipartite entangled states \cite{CDS08,YCGD10,CDS10,CCDJW10,YGD14,VC15,VC17,GMO20}. It is known that the tensor rank is not multiplicative under the Kronecker product. This is the reason why multicopy entanglement transformation by SLOCC is quite challenging. In Refs. \cite{CDS08,CCDJW10}, the tensor rank of two copies of the three-qubit $\W$ state is shown to be $\rk(\mathcal{W}_3\boxtimes\mathcal{W}_3)=7$, and in general the tensor rank of two copies of the $n$-qubit $\W$ state is shown to be $\rk(\mathcal{W}_n\boxtimes\mathcal{W}_n)=3n-2$. The tensor rank has also been shown to not be multiplicative under the tensor product \cite{CJZ18}. In Ref. \cite{CF18} it is shown that $\rk(\mathcal{W}_3\otimes\mathcal{W}_3)=8$ but the tensor rank of the tensor product of two $n$-qubit $\W$ states is still unknown.

In the following, we present the definitions of the SLOCC and asymptotic SLOCC transformations that are, respectively, known as restriction and degeneration in algebraic geometry and algebraic complexity theory.

\begin{definition}\label{def:SLOCC-transformation}
Let $|\psi\rangle\in U_1\otimes\cdots\otimes U_n$ and $|\varphi\rangle\in V_1\otimes\cdots\otimes V_n$ be two $n$-partite quantum states, where $U_i$ and $V_i$ are the Hilbert spaces of individual subsystems. We say that $|\psi\rangle$ can be transformed into $|\varphi\rangle$ via SLOCC (denoted by $|\psi\rangle\xrightarrow[]{\text{SLOCC}}|\varphi\rangle$) if there exist linear maps $A_i:U_i\to V_i$ such that
\begin{equation}\label{SLOCC}
(\otimes_{i=1}^{n}A_i)|\psi\rangle=|\varphi\rangle\,.
\end{equation}
\end{definition}

A generalization of the concept of SLOCC conversion is that of asymptotic SLOCC conversion. Here, instead of an exact transformation according to Eq. \eqref{SLOCC}, we consider asymptotic transformations between quantum states by local operations.

\begin{definition}\label{def:degeneration}
Let $|\psi\rangle\in U_1\otimes\cdots\otimes U_n$ and $|\varphi\rangle\in V_1\otimes\cdots\otimes V_n$ be two $n$-partite quantum states, where $U_i$ and $V_i$ are the Hilbert spaces of individual subsystems. We say that $|\psi\rangle$ degenerates into $|\varphi\rangle$ with error degree $e$ via SLOCC (denoted by $|\psi\rangle\xdashrightarrow[]{\text{SLOCC}}|\varphi\rangle$) if there exist linear maps $A_i(\varepsilon):U_i\to V_i$ depending polynomially on $\varepsilon$ such that
\begin{equation}\label{degenration}
(\otimes_{i=1}^{n}A_i(\varepsilon))|\psi\rangle=\varepsilon^d|\varphi\rangle+\sum_{l=1}^e\varepsilon^{d+l}|\tilde{\varphi}_l\rangle\,,
\end{equation}
for some state $|\tilde{\varphi}_l\rangle$ and $d\in\mathbbm{N}$ which is called the approximation degree.
\end{definition}
Indeed, if the quantum state $|\psi\rangle$ degenerates into the quantum state $|\varphi\rangle$, then $|\varphi\rangle$ can be approximated to arbitrary precision by restrictions of $|\psi\rangle$, i.e.,
\begin{equation}\label{asymptoticSLOCC}
\lim_{\varepsilon\to 0}\frac{1}{\varepsilon^d}(\otimes_{i=1}^{n}A_i(\varepsilon))|\psi\rangle=|\varphi\rangle\,.
\end{equation}

In a similar spirit to the LOCC-based entanglement dilution \cite{Nielsen-Chuang}, we can use a quantity that indicates the minimum number of copies of a source quantum state $|\psi\rangle$ that can be used to obtain a single copy of the target quantum state $|\varphi\rangle$ by SLOCC transformation, in an asymptotic setting. This quantity is the rate of asymptotic SLOCC transformation from $|\psi\rangle$ into $|\varphi\rangle$ and is defined as follows
\begin{equation}\label{rate}
\omega(\psi,\varphi)=\lim_{n\to\infty}\frac{1}{n}\inf\left\{m\in\mathbbm{N}\big||\psi\rangle^{\boxtimes m}\xrightarrow[]{\text{SLOCC}}|\varphi\rangle^{\boxtimes n}\right\}\,.
\end{equation}


\subsection{Concise tensors}
Informally, a tensor is concise if it cannot be written as a tensor in a smaller ambient space.
For example, a tensor $\mathcal{T}\in\mathbbm{C}^{a}\otimes\mathbbm{C}^{b}\otimes\mathbbm{C}^{c}$ is concise if its multilinear rank is $(a,b,c)$, which means that the tensor $\mathcal{T}$ uses all dimensions of the local spaces. In the following, we define concise tensors for multipartite systems.

\begin{definition}\label{def:concise}
A tensor $\mathcal{T}\in V_1\otimes\cdots\otimes V_n$ is called \emph{concise in the first factor}, or \emph{$1$-concise}, if $\mathcal{T}\notin V'_1\otimes V_2\otimes\cdots\otimes V_n$ with $V'_1\subsetneq V_1$. Conciseness in other factors is defined analogously. A tensor is called \emph{concise} if it is $i$-concise for all $i\in\{1,\ldots,n\}$.
\end{definition}

\begin{remark}
In quantum information theory, concise tensors correspond to quantum states with maximal one-to-group marginal entanglement (bipartite entanglement between each single party and the remaining parties). Conciseness can be alternatively characterized as all single-party reduced density matrices being full rank.
\end{remark}

The following lemma gives several equivalent characterizations of $1$-concise tensors.

\begin{lemma}\label{lem:concise-tfae}
Let $\mathcal{T}\in V_1\otimes\cdots\otimes V_n$ be a tensor and $\dim V_i=d_i$.
The following statements are equivalent:
\begin{enumerate}
    \item $\mathcal{T}$ is $1$-concise;
    \item For every nonzero covector $\langle f| \in V_1^{\vee}$ the contraction $\langle f| \mathcal{T}$ is nonzero;
    \item For every basis $\{|e_j\rangle\mid j\in\mathbbm{Z}_{d_1}\}$ of $V_1$ the decomposition $\mathcal{T} = \sum_{j = 0}^{d_1-1} |e_j\rangle\otimes\mathcal{T}_j$ has all $\mathcal{T}_j$ nonzero.
\end{enumerate}
\end{lemma}
\begin{proof}
$(1) \Leftrightarrow (2)$: 
Note that $\langle f| \mathcal{T} = 0$ iff $\mathcal{T}$ is in $(\ker\langle f|)\otimes V_2\otimes\cdots\otimes V_n$.
If $\mathcal{T}$ is $1$-concise, then $\langle f|\mathcal{T}=0$ iff $\ker\langle f|=V_1$, that is, $\langle f|=0$.
Conversely, assume that $\mathcal{T}$ is not $1$-concise, that is, $\mathcal{T} \in V_1'\otimes V_2 \otimes\cdots\otimes V_n$ with $V_1'\subsetneq V_1$.
There exists a nonzero covector $\langle f|$ vanishing on $V_1'$, and for this covector we have $\langle f|\mathcal{T}=0$.

$(2) \Rightarrow (3)$: Let $\{\langle f_j|\mid j\in\mathbbm{Z}_{d_1}\}$ be the dual basis. $\mathcal{T}_j = \langle f_{j}| \mathcal{T}$ is nonzero.

$(3) \Rightarrow (2)$: For every nonzero $\langle f|$ there exists a basis $\{|e_j\rangle\mid j\in\mathbbm{Z}_{d_1}\}$ such that $\langle f|e_{j}\rangle=0$ for $j>0$ and $\langle f|e_0\rangle=1$. We have $\langle f|\mathcal{T}=\mathcal{T}_0\neq0$.
\end{proof}

\begin{corollary}\label{cor:concise-matrix}
A tensor $\mathcal{T}\in V_1\otimes V_2$ is $1$-concise iff $\rk(\mathcal{T})=\dim V_1$.
\end{corollary}

We will need the following property of rank decompositions for $i$-concise tensors.
\begin{lemma}\label{lem:concise-rank}
Let $\mathcal{T}\in V_1\otimes\cdots\otimes V_n$ be an $i$-concise tensor. If
\begin{equation}\label{rank-decomposition}
\mathcal{T}= \sum_{p = 1}^r v_1^{(p)}\otimes\cdots\otimes v_n^{(p)}\,,
\end{equation}
is a tensor rank decomposition of $\mathcal{T}$, then the vectors $\{v_i^{(1)},\ldots,v_i^{(r)}\}$ span $V_i$.
\end{lemma}
{\it Proof.}
We prove the statement for $1$-concise tensors. Let $U=\Span\{v_1^{(1)},\ldots, v_1^{(r)}\}$. Note that all summands of rank decomposition are contained in $U\otimes V_2\otimes\cdots\otimes V_n$. It follows that $\mathcal{T}$ also lies in this space. Since $\mathcal{T}$ is a $1$-concise tensor, we then have $U=V_1$.

The proof for $i$-concise tensors is analogous.
\qed


\section{Persistent Tensors}\label{Sec.iii}

The substitution method is a method to obtain lower bounds for the tensor rank by zeroing the summands in a rank decomposition by applying appropriate projection maps to the tensor factors (see \cite[Ch.17]{BCS97} or \cite[Appx.~B]{AFT11}). A lower bound is obtained by keeping track of the number of summands zeroed. In this section, we introduce a class of tensors for which we can prove the tensor rank lower bounds by repeated application of the substitution method.
We call these tensors persistent.
For persistent tensors in $\otimes^n\mathbbm{C}^d$ we get a lower bound of $(n-1)(d-1)+1$.

\begin{definition}\label{def:persistent-tensor}
We define {\bf\emph{persistent tensors}} inductively.
\begin{itemize}
\item[(i)] A tensor $\mathcal{P}\in V_1\otimes V_2$ is persistent if it is $1$-concise.
\item[(ii)] A tensor $\mathcal{P}\in V_1\otimes\cdots\otimes V_n$ with $n > 2$ is persistent if it is $1$-concise and there exists
a subspace $S \subsetneq V_1^{\vee}$ such that the contraction $\langle f|\mathcal{P}\in V_2\otimes\cdots\otimes V_n$ is persistent whenever $\langle f|\notin S$.
\end{itemize}
\end{definition}

The following lemma gives different characterizations of the class of persistent tensors which are useful for checking persistence.
\begin{lemma}\label{lem:persistence-equivalent}
Let $\mathcal{P}\in V_1\otimes\cdots\otimes V_n$ ($n>2$) be a persistent tensor with $\dim V_i=d_i$. The following statements are equivalent:
\begin{enumerate}
    \item $\mathcal{P}$ is persistent.
    \item $\mathcal{P}$ is $1$-concise and there exists a nonzero vector $|e\rangle\in V_1$ such that the following implication holds:
    \[
    \langle f|e\rangle\neq0 \Rightarrow \text{$\langle f|\mathcal{P}$ is persistent.}
    \]
    \item $\mathcal{P}$ is $1$-concise and there exists a nonzero vector $|e\rangle \in V_1$ such that the following implication holds:
    \[
    \langle f|e\rangle=1 \Rightarrow \text{$\langle f|\mathcal{P}$ is persistent.}
    \]
    \item For every basis $\{|e_j\rangle\mid j\in\mathbbm{Z}_{d_1}\}$ of $V_1$ the decomposition
    \begin{equation}\label{eq:basis-decomposition-persistent}
    \mathcal{P}=\sum_{j=0}^{d_1-1}|e_j\rangle\otimes\mathcal{P}_j\,,
    \end{equation}
    has all $\mathcal{P}_j$ nonzero and at least one of them is persistent.
\end{enumerate}
\end{lemma}
\begin{proof}
$(1) \Rightarrow (2)$:
Let $S \subsetneq V_1^{\vee}$ be the subspace in the definition of persistence. Choose any nonzero $|e\rangle \in S^{\perp}$.

$(2) \Rightarrow (1)$:
Take $S=|e\rangle^{\perp}$.

$(2) \Rightarrow (3)$: Trivial.

$(3) \Rightarrow (2)$: If $\langle f|e\rangle=a\neq 0$, then $\langle f'|e\rangle=1$ for $\langle f'|=\frac{1}{a}\langle f|$. Hence $\frac{1}{a}\langle f|\mathcal{P}$ is persistent, and $\langle f|\mathcal{P}$ is persistent because persistence is scaling-invariant.

$(1) \Rightarrow (4)$:
Because $\mathcal{P}$ is $1$-concise, all $\mathcal{P}_i$ are nonzero. Let $\{\langle f_j|\mid j\in\mathbbm{Z}_{d_1}\}$ be the dual basis to $\{|e_j\rangle\mid j\in\mathbbm{Z}_{d_1}\}$. At least one $\langle f_j|$ does not lie in the subspace $S\subsetneq V_1^{\vee}$ from the definition of persistence. It follows that $\mathcal{P}_j=\langle f_j|\mathcal{P}$ is persistent.

$(4) \Rightarrow (3)$:
Let $\{|e_j\rangle\mid j\in\mathbbm{Z}_{d_1}\}$ be a basis such that the decomposition~\eqref{eq:basis-decomposition-persistent} has the minimum possible number of persistent tensors $\mathcal{P}_j$.
Assume without loss of generality that $\mathcal{P}_0$ is persistent.

For every $\alpha_1,\ldots,\alpha_{d_1-1}$ we can rewrite the decomposition~\eqref{eq:basis-decomposition-persistent} to get
\begin{equation}
\mathcal{P}=|e_0\rangle\otimes\big(\mathcal{P}_0+\sum_{j=1}^{d_1-1} \alpha_j\mathcal{P}_j\big)+\sum_{j=1}^{d_1-1}(|e_j\rangle-\alpha_j|e_0\rangle)\otimes\mathcal{P}_j.
\end{equation}
This is a decomposition corresponding to a different basis $|e_0'\rangle=\frac{1}{\alpha_0}|e_0\rangle, |e_j'\rangle=|e_j\rangle-\frac{\alpha_j}{\alpha_0}|e_0\rangle$.
Since the number of persistent slices in this decomposition cannot be less than that in the original, the tensor $\mathcal{P}_0+\sum_{j=1}^{d_1-1}\alpha_j\mathcal{P}_j$ is persistent.

Let $\langle f|\in V_1^{\vee}$ be a covector such that $\langle f|e_0\rangle=1$.
Note that $\langle f|\mathcal{P}=\mathcal{P}_0+\sum_{j=1}^{d_1-1}\langle f|e_j\rangle\mathcal{P}_j$ is persistent by the previous discussion.
We have proven $(3)$ with $|e\rangle=|e_0\rangle$.
\end{proof}

In the following, we present some examples of persistent and non-persistent tensors.

\begin{itemize}
\item[(i).] Non-persistent tensors:
\begin{itemize}
\item[1.] The diagonal tensor $\mathcal{G}(d,n)$ (correspondingly, $n$-qudit $\GHZ$ state) is not a persistent tensor for $n>2$ and $d\geq{2}$. This can be understood using \Cref{lem:persistence-equivalent}(4). We have
\begin{equation}
\mathcal{G}(d,n)=\sum_{j=0}^{d - 1}|j\rangle\otimes\mathcal{T}_j\quad\text{and}\quad\mathcal{T}_j=|j\rangle^{\otimes(n-1)}\,,
\end{equation}
where all $\mathcal{T}_j$ are not $1$-concise and therefore not persistent.
\item[2.] The Dicke state $\mathcal{D}_4^2=|0011\rangle+|0101\rangle+|0110\rangle+|1001\rangle+|1010\rangle+|1100\rangle$ is not a persistent tensor. This can be seen from the decomposition~\eqref{eq:basis-decomposition-persistent} corresponding to the basis $|\pm\rangle=|0\rangle\pm|1\rangle$ and the fact that $\mathcal{W}_3\pm\overline{\mathcal{W}}_3\equiv\mathcal{G}(2,3)$, where $\overline{\mathcal{W}}_3=|011\rangle+|101\rangle+|110\rangle$.
\item[3.] All unnormalized multiqubit Dicke states
\begin{equation}\label{Dicke-qubit}
\mathcal{D}_n^l=\sum_{\mathfrak{p}\in\mathfrak{S}_n}\mathfrak{p}\big\{|0\rangle^{\otimes(n-l)}\otimes|1\rangle^{\otimes l}\big\}\,,
\end{equation}
with $l$ excitations are not persistent tensors except when $l=1$. This can be understood from the previous example.
\end{itemize}
\item[(ii).] Persistent tensors:
\begin{itemize}
\item[1.] $\mathcal{W}_n$ is a persistent tensor because for every $\langle f|$ such that $\langle f|0\rangle=1$ we have
\begin{equation}
\langle f|\mathcal{W}_n=\mathcal{W}_{n-1}+\langle f|1\rangle|0\rangle^{\otimes(n-1)}\,,
\end{equation}
which is equivalent to $\mathcal{W}_{n-1}$. Repeating this construction, we arrive at the base case $\mathcal{W}_2=|01\rangle+|10\rangle$ which is a persistent tensor. Indeed, the $n$-qubit $\W$ state is the only symmetric persistent tensor in multiqubit systems.
\item[2.] An example of a nonsymmetric persistent tensor is the four-qubit state $\mathcal{T} = \alpha^2|0011\rangle+\beta^2|0101\rangle+(\alpha \pm \beta)^2|0110\rangle+|1001\rangle+|1010\rangle+|1100\rangle$. For every $\langle f|$ such that $\langle f|1\rangle = 1$ the contraction $\langle f|\mathcal{T}$ is equivalent to $\mathcal{W}_3$. This can be checked by computing the tangle~\cite{LLHL06}.
\item[3.] As another example, 3-qutrit $\mathcal{Y}_3=|002\rangle+|020\rangle+|200\rangle+|011\rangle+|101\rangle+|110\rangle$ is a persistent tensor. In the following, we will show that the $n$-qutrit $\Y$ state given by
\begin{equation}\label{Y}
\mathcal{Y}_n=\sum_{\mathfrak{p}\in\mathfrak{S}_n}\mathfrak{p}\{|0\rangle^{\otimes(n-2)}(|02\rangle+|11\rangle)\}\,,
\end{equation}
which corresponds to a symmetric tensor in $\otimes^n\mathbbm{C}^3$, is a persistent tensor.
\end{itemize}
\end{itemize}

\begin{theorem}\label{theo:PTLB}
If $\mathcal{P}\in V_1\otimes\cdots\otimes V_n$ is a persistent tensor and $d_k=\dim{V_k}$, then
\begin{equation}\label{PTLB}
\rk(\mathcal{P})\geq\sum_{k=1}^{n-1}(d_k-1) + 1\,.
\end{equation}
Moreover, in every rank decomposition
\begin{equation}\label{P-rank-decomposition}
\mathcal{P}=\sum_{p=1}^r u_1^{(p)}\otimes\cdots\otimes u_n^{(p)}\,,
\end{equation}
one can permute the summands in such a way that the rearranged decomposition
\begin{equation}\label{P-rank-decomposition-2}
\mathcal{P}=\sum_{p=1}^r v_1^{(p)}\otimes\cdots\otimes v_n^{(p)}\,,
\end{equation}
has the following property:
for every $j<n$ the vectors $\{v_j^{(D_j+1)},\ldots, v_j^{(D_j + d_j)}\}$ form a basis of $V_j$, where $D_j = \sum_{k = 1}^{j - 1} (d_k - 1)$ (for $j = 1$ we take $D_1 = 0$).
\end{theorem}
{\it Proof.}
We prove the statement by induction.

Base case: $n = 2$. If $n = 2$, then $\mathcal{P}$ is persistent iff it is $1$-concise. By \Cref{cor:concise-matrix} we have $\rk(\mathcal{P}) = d_1$, so the required lower bound is maintained.
Moreover, in any decomposition
\begin{equation}
\mathcal{P}=\sum_{p=1}^r u_1^{(p)}\otimes u_2^{(p)}\,,
\end{equation}
Regarding \Cref{lem:concise-rank}, $\Span\{u_1^{(1)},\ldots,u_1^{(r)}\}=V_1$. Therefore, we can permute the summands to get a decomposition
\begin{equation}
\mathcal{P}=\sum_{p=1}^r v_1^{(p)}\otimes v_2^{(p)}\,,
\end{equation}
where $\{v_1^{(1)},\ldots,v_1^{(d_1)}\}$ form a basis of $V_1$.

Consider now the case $n > 2$. Let $\mathcal{P}$ be a persistent tensor in $V_1 \otimes \cdots \otimes V_n$ and let Eq. \eqref{P-rank-decomposition} be a rank decomposition of $\mathcal{P}$. We rearrange the summands of this decomposition in several steps.

First, since $\mathcal{P}$ is $1$-concise, based on \Cref{lem:concise-rank}, $V_1=\Span\{u_1^{(1)}, \ldots, u_1^{(r)}\}$. So, we can permute the summands to obtain a rearranged decomposition in Eq. \eqref{P-rank-decomposition-2} such that $\{v_1^{(1)},\ldots,v_1^{(d_1)}\}$ form a basis of $V_1$. We choose the order of this basis in such a way that in the decomposition
\begin{equation}
\mathcal{P} = \sum_{k = 1}^{d_1} v_1^{(k)} \otimes \mathcal{P}_k\,,
\end{equation}
the tensor $\mathcal{P}_{d_1}$ is persistent.

Let $V_1'=\Span\{v_1^{(1)}, \ldots, v_1^{(d_1 - 1)}\}$.
As a second step, we separate the summands with $v_1^{(p)} \notin V_1'$ from those with $v_1^{(p)} \in V_1'$.
We rearrange the summands with indices from $d_1 + 1$ to $r$ to get a second rearranged decomposition
\begin{equation}\label{P-rank-decomposition-3}
\mathcal{P}=\sum_{p=1}^r w_1^{(p)}\otimes\cdots\otimes w_n^{(p)}\,,
\end{equation}
such that $w_1^{(p)}\notin V_1'$ if $d_1\leq p\leq s$ and $w_1^{(p)}\in V_1'$ if $p > s$ for an appropriate $s \leq r$.

Let $\langle f|$ be a covector such that $\langle f|v_1^{(k)}\rangle=0$ if $k<d_1$ and $\langle f|v_1^{(d_1)}\rangle=1$.
We have
\begin{equation}
\mathcal{P}_{d_1} =\langle f|\mathcal{P}=\sum_{p=d_1}^s \langle f|w_1^{(p)}\rangle w_2^{(p)}\otimes\cdots\otimes w_n^{(p)}\,,
\end{equation}
which is a rank decomposition for $\mathcal{P}_{d_1}$.

By the induction hypothesis, the number of summands in this decomposition, $s-d_1+1$, is at least $\sum_{k=2}^{n-1}(d_k-1)+1$, from which we obtain $r \geq s \geq \sum_{k = 1}^{n - 1} (d_k - 1) + 1$ as required.
Moreover, we can rearrange the summands to get the following
\begin{equation}
\mathcal{P}_{d_1}=\langle f|\mathcal{P} =\sum_{p = d_1}^s\langle f|y_1^{(p)}\rangle y_2^{(p)}\otimes\cdots\otimes y_n^{(p)},
\end{equation}
with $\{y_j^{(D_j + 1)},\ldots, y_j^{(D_j + d_j)}\}$ being a basis of $V_j$.
Applying the same permutation to the summands with indices from $d_1$ to $s$ in Eq. \eqref{P-rank-decomposition}, we get
\begin{equation}
\mathcal{P}=\sum_{p=1}^{r} y_1^{(p)}\otimes\cdots\otimes y_n^{(p)}\,.
\end{equation}
Note that $\{y_1^{(1)},\ldots, y_1^{(d_1 - 1)}, y_1^{(d_1)}\}$ is still a basis of $V_1$ since $y_1^{(d_1)} \notin V_1' = \Span\{y_1^{(1)},\ldots, y_1^{(d_1 - 1)}\}$.
\qed

An alternative proof of \Cref{theo:PTLB} can be found in \Cref{AppxA}.

\begin{corollary}\label{cor:W-rank}
The tensor rank of the $n$-qubit $\W$ state is $n$.
\end{corollary}
{\it Proof.}
The upper bound is obvious from the definition of $\mathcal{W}_n$, which has $n$ summands. According to \Cref{theo:PTLB}, the lower bound of the tensor rank of $\mathcal{W}_n$ is $n$ as it is a persistent tensor.
\qed


\section{Multiqudit generalization of $\rm{W}$ state}
We now introduce several families of multipartite tensors in $\otimes^n\mathbbm{C}^d$ (corresponding to $n$-qudit states) which can be thought as different generalizations of multiqubit $\W$ states within multiqudit systems. In the tripartite case, these tensors have been studied before in the context of matrix multiplication complexity in connection with the Coppersmith-Winograd algorithm~\cite{CW90}.

The first family we present we call $n$-qudit $\LL$ states
\begin{equation}\label{L-state}
\mathcal{L}(d,n)=\sum_{j_1+\cdots+j_n=d-1}|j_1\cdots j_n\rangle\,.
\end{equation}
These states are a special case of the weight states considered by Christandl et al. in Ref. \cite{CGFW21}.
For $n=d=3$ this tensor appeared as the $\Y$ state in Ref. \cite{GM21}.
In algebraic complexity theory, a tensor equivalent to $\mathcal{L}(d, 3)$ appeared as the structure tensor of truncated polynomial multiplication or as a ``lower triangular'' version of the cyclic group tensor~\cite{AVW18}.

The second family we introduce is the family of $n$-qudit $\M$ states, which generalizes the Coppersmith-Winograd tensors used in matrix multiplication algorithms~\cite{CW90} to the multipartite case.
We give two versions of these tensors, which are SLOCC equivalent.
The first version is
\begin{equation}\label{M-state}
\begin{split}
\mathcal{M}(d,n)&=\sum_{\mathfrak{p}\in\mathfrak{S}_n}\mathfrak{p}\Big\{\sum_{j=0}^{\lfloor\frac{d-1}{2}\rfloor}|0\rangle^{\otimes(n-2)}|j\rangle|d-j-1\rangle\Big\} \\
&=\sum_{i=0}^{n-1}|0\rangle^{\otimes(n-i-1)}|d-1\rangle|0\rangle^{\otimes i}+\sum_{i+k+l=n-2}\sum_{j=1}^{d-2}|0\rangle^{\otimes i}|j\rangle|0\rangle^{\otimes k}|d-j-1\rangle|0\rangle^{\otimes l}\,.
\end{split}
\end{equation}
The second version is (for $d\geq3$)
\begin{equation}\label{M-state-2}
\begin{split}
\mathcal{M}'(d,n)&=\sum_{\mathfrak{p}\in\mathfrak{S}_n}\mathfrak{p}\Big\{|0\rangle^{\otimes(n-1)}|d-1\rangle+\sum_{j=1}^{d-2}|0\rangle^{\otimes(n-2)}|jj\rangle\Big\} \\ 
&=\sum_{i=0}^{n-1}|0\rangle^{\otimes(n-i-1)}|d-1\rangle|0\rangle^{\otimes i}+\sum_{i+k+l=n-2}\sum_{j=1}^{d-2}|0\rangle^{\otimes i}|j\rangle|0\rangle^{\otimes k}|j\rangle|0\rangle^{\otimes l}\,.
\end{split}
\end{equation}
For $d=3$, the two versions are equal. For $d\ge4$, the two versions are SLOCC equivalent, because $\mathcal{M}$ is transformed into $\mathcal{M}'$ by applying to each factor the following change of basis
\begin{equation}
\begin{cases}
|j\rangle\mapsto\frac{1}{\sqrt{2}}(|j\rangle+\mathbf{i}|d-j-1\rangle) \\
|d-j-1\rangle\mapsto\frac{1}{\sqrt{2}}(|j\rangle-\mathbf{i}|d-j-1\rangle)
\end{cases} \text{for} \quad 1\leq j\leq\lfloor\frac{d-2}{2}\rfloor\,,
\end{equation}
where $\mathbf{i}=\sqrt{-1}$. In fact, it is a direct consequence of the Schmidt decomposition that $|j\rangle|d-j-1\rangle+|d-j-1\rangle|j\rangle$ is equivalent to $|j\rangle|j\rangle+|d-j-1\rangle|d-j-1\rangle$ \cite{Schmidt}.

The last family we present is the family of $n$-qudit $\N$ states defined as
\begin{equation}\label{N-state}
\mathcal{N}(d,n)=|0\rangle^{\otimes(n-1)}|d-1\rangle+\sum_{i=0}^{n-2}\sum_{j=1}^{d-1}|0\rangle^{\otimes(n-i-2)}|j\rangle|0\rangle^{\otimes i}|d-j-1\rangle\,.
\end{equation}
By applying the map $|j\rangle\mapsto|d-j-1\rangle$ to the last tensor factor, we get an equivalent tensor as follows
\begin{equation}\label{N-state-2}
\mathcal{N}'(d,n)=|0\rangle^{\otimes n}+\sum_{i=0}^{n-2}\sum_{j=1}^{d-1}|0\rangle^{\otimes(n-i-2)}|j\rangle|0\rangle^{\otimes i}|j\rangle\,.
\end{equation}
This form of $n$-qudit $\N$ state generalizes the tripartite tensor considered by Copersmith-Winograd as the asymmetric version of its construction~\cite{CW90}.

All three families of tensors generalize the $n$-qubit $\W$ state in the sense that $\mathcal{W}_n=\mathcal{L}(2, n)=\mathcal{M}(2, n)=\mathcal{N}(2, n)$. Moreover, it is easy to see that $\mathcal{W}_n^{\boxtimes 2}$ is equivalent to $\mathcal{M}(4, n)$ under the identification $|00\rangle \mapsto |0\rangle, |01\rangle \mapsto |1\rangle, |10\rangle \mapsto |2\rangle, |11\rangle \mapsto |3\rangle$.

All these families consist of persistent tensors, which follows from a more general statement below.

\begin{theorem}
Let $\mathcal{T}\in\otimes^n\mathbbm{C}^d$ be a tensor of the form
\begin{equation}
\mathcal{T}=\sum_{j_1+\cdots+j_n< d} t_{j_1\cdots j_n} |j_1\cdots j_n\rangle\,.
\end{equation}
If the coefficients before $|0\rangle^{\otimes(n-i-2)}|j\rangle|0\rangle^{\otimes i}|d-j-1\rangle$ are nonzero for all $i\leq n-2$ and $j\leq d-1$ then $\mathcal{T}$ is a persistent tensor.
\end{theorem}
\begin{proof}
We prove the statement by induction on $n$.

If $n=2$, then 
\begin{equation}
\mathcal{T} =\sum_{j=0}^{d-1} t_{j,d-j-1}|j\rangle\otimes\Big(|d-j-1\rangle+\sum_{k = 0}^{d-j-2} \frac{t_{jk}}{t_{j,d-j-1}} |k\rangle\Big)\,,
\end{equation}
has matrix rank $d$ and therefore is $1$-concise.

For $n>2$, note that for $\langle f|=\sum_{j=0}^{d-1} f_j\langle j|$ we have
\begin{equation}
\langle f|\mathcal{T}=\sum_{j_2+\cdots+j_n< d} s_{j_2\cdots j_n} |j_2 \cdots j_n\rangle\quad\text{where}\quad s_{j_2\cdots j_n}=\sum_{j_1=0}^{d-(j_2+\cdots+ j_n)-1} f_{j_1} t_{j_1\cdots j_n}\,.
\end{equation}

If $\langle f|\neq 0$ and $j$ is the minimum index such that $f_j \neq 0$, then $s_{0\cdots0,d-j-1}= f_j t_{j0\cdots0,d-j-1}$ $\neq0$, so $\langle f|\mathcal{T} \neq 0$. By \Cref{lem:concise-tfae} $\mathcal{T}$ is $1$-concise. Additionally, if $\langle f|0\rangle= 1$ and $j_2+\cdots+j_n=d-1$, then $s_{j_2\cdots j_n}=t_{0j_2\cdots j_n}$. In particular, the coefficients before $|0\rangle^{\otimes (n-i-3)} |j\rangle |0\rangle^{\otimes i} |d-j-1\rangle$ in $\langle f|\mathcal{T}$ are nonzero, so by the induction hypothesis $\langle f|\mathcal{T}$ is persistent.
Therefore, $\mathcal{T}$ is persistent by \Cref{lem:persistence-equivalent}(3) with $|e\rangle=|0\rangle$.
\end{proof}

\begin{corollary}
The tensors $\mathcal{L}(d,n)$, $\mathcal{M}(d,n)$ and $\mathcal{N}(d,n)$ are persistent.
\end{corollary}

The persistence allows us to use the lower bound of~\Cref{theo:PTLB} to find the ranks of $\mathcal{L}(d, n)$, $\mathcal{M}(d, n)$ and $\mathcal{N}(d, n)$.

\begin{theorem}
The tensor rank of the $n$-qudit $\LL$ state is $\rk(\mathcal{L}(d,n))=(n-1)(d-1)+1$.
\end{theorem}
\begin{proof}
The lower bound follows from~\Cref{theo:PTLB}.
For the upper bound, we give an explicit decomposition.

Let $r=(n-1)(d-1)+1$ and let $\zeta =\exp(\frac{2\pi\mathbf{i}}{r})$ be a primitive $r$-th root of unity.
Using the property of roots of unity
\begin{equation}\label{roots-unity-2}
\sum_{p = 0}^{r-1} \zeta^{p q}=
\begin{cases}
r & \text{if $r\,|\,q$} \\
0 & \text{otherwise,}
\end{cases}
\end{equation}
we see that
\begin{equation}
\begin{split}
\mathcal{L}(d,n)&=\sum_{j_1+\cdots+j_n=d-1}|j_1\cdots j_n\rangle=\frac{1}{r}\sum_{s=0}^{n(d-1)}\sum_{j_1+\cdots+j_n=s} \Big(\sum_{p=0}^{r-1}\zeta^{p(s-d+1)}\Big)|j_1\cdots j_n\rangle
\\
&=\frac{1}{r}\sum_{j_1+\cdots+j_n=s}\Big(\sum_{p=0}^{r-1}\zeta^{p(\sum_{i=1}^n j_i-d+1)}\Big)|j_1\cdots j_n\rangle=\frac{1}{r}\sum_{p=0}^{r-1}\zeta^{p(-d+1)}\Big(\sum_{j=0}^{d-1}\zeta^{p j}|j\rangle\Big)^{\otimes n}\,.
\end{split}
\end{equation}
\end{proof}

\begin{corollary}\label{cor:tensor-rank-Y}
The tensor rank of the $n$-qutrit $\Y$ state $\mathcal{Y}_n = \mathcal{L}(3, n)$ is $2n-1$.
\end{corollary}

\begin{theorem}\label{lem:Tensor-rank-M}
The tensor rank of the $n$-qudit $\M$ state is $\rk(\mathcal{M}(d,n))=(n-1)(d-1)+1$.
\end{theorem}
\begin{proof}
For the lower bound, we again use~\Cref{theo:PTLB}.

We prove the upper bound for the SLOCC equivalent tensor $\mathcal{M}'(d, n)$.
Note that $\mathcal{M}'(d,n)$ is the sum of $d - 2$ tensors of the form $\sum_{\mathfrak{p} \in \mathfrak{S}_n} \mathfrak{p}\lbrace|0\rangle^{\otimes(n-2)}|jj\rangle\rbrace$, which are equivalent to the Dicke state $\mathcal{D}^2_n$, and the tensor $\sum_{\mathfrak{p}\in\mathfrak{S}_n} \mathfrak{p}\lbrace|0\rangle^{\otimes(n-1)}|d-1\rangle\rbrace$, which is equivalent to $\mathcal{W}_n$.
The rank of $\mathcal{D}^2_n$ is $n-1$ \cite{CDS10} and the rank of $\mathcal{W}_n$ is $n$. We obtain the required upper bound by summing these ranks.
\end{proof}

\begin{corollary}\label{cor:tensor-rank-WW}
$\rk(\mathcal{W}_n\boxtimes\mathcal{W}_n)=\rk(\mathcal{M}(4,n))=3n-2$ which already has been obtained in Ref. \cite{CCDJW10}.
\end{corollary}

\begin{theorem}
The tensor rank of the $n$-qudit $\N$ state is $\rk(\mathcal{N}(d,n))=(n-1)(d-1)+1$.
\end{theorem}
\begin{proof}
The lower bound again follows from~\Cref{theo:PTLB}, and the upper bound is obvious from the definition of $\mathcal{N}$, which has $(n-1)(d-1)+1$ summands.
\end{proof}

Thus, the three families of persistent tensors we introduced have rank $(n-1)(d-1)+1$, which is the minimum possible rank for a persistent tensor in $\otimes^n\mathbbm{C}^d$.
We can show that their border rank also has the minimum possible value $d$.

\begin{theorem}\label{theo:LMN-degenerations}
For $n\geq3$ we have a chain of degenerations
\begin{equation}\label{LMN-degenerations}
\mathcal{L}(d,n)\xdashrightarrow[]{\text{SLOCC}}\mathcal{M}(d,n)\xdashrightarrow[]{\text{SLOCC}}\mathcal{N}(d,n)\,.
\end{equation}
\end{theorem}
\begin{proof}
To degenerate from from $\mathcal{L}(d, n)$ to $\mathcal{M}(d, n)$, one can apply the family of linear maps $A(\varepsilon)=\mathrm{diag}(\varepsilon^{-2},\varepsilon^{n-2},\ldots,\varepsilon^{n-2},\varepsilon^{2(n-1)})$ to each tensor factor and let $\varepsilon\to0$.

To degenerate from $\mathcal{M}(d, n)$ to $\mathcal{N}(d, n)$, apply $A(\varepsilon)=\mathrm{diag}(1,\varepsilon,\ldots,\varepsilon,1)$ to the first $n-1$ factors and $A(\varepsilon)^{-1}$ to the last factor, and let $\varepsilon\to 0$.
\end{proof}

\begin{theorem}\label{theo:LMN-brank}
$\brk(\mathcal{L}(d,n)) = \brk(\mathcal{M}(d,n)) = \brk(\mathcal{N}(d,n)) = d$.
\end{theorem}
\begin{proof}
The lower bound follows from the $1$-conciseness of the tensors.

We give an explicit approximation for $\mathcal{L}(d, n)$.
Let $\xi=\exp(\frac{2\pi\mathbf{i}}{d})$ be the primitive $d$-th root of unity.
Using the property of roots of unity as follows
\begin{equation}\label{roots-unity}
\sum_{p = 0}^{d-1} \xi^{p q}=
\begin{cases} d & \text{if $d\,|\,q$} \\
0 & \text{otherwise,}
\end{cases}
\end{equation}
we can give an approximation of the $n$-qudit $\LL$ state as follows
\begin{equation}\label{border-rank-L}
\mathcal{L}(d,n)=\lim_{\varepsilon\to 0} \frac{1}{d\,\varepsilon^{d-1}}\sum_{p = 0}^{d-1} \xi^p \Big(\sum_{j=0}^{d-1} \varepsilon^j\xi^{p j}|j\rangle\Big)^{\otimes n}\,.
\end{equation}
The upper bound for $\mathcal{M}(d,n)$ and $\mathcal{N}(d,n)$ can be transferred from $\mathcal{L}(d,n)$ using degenerations from~\Cref{theo:LMN-degenerations}.
\end{proof}

Alternatively, we can give approximate decompositions for $\mathcal{M}'(d, n)$ and $\mathcal{N}'(d, n)$ as follows
\begin{equation}\label{M-state-BR}
\mathcal{M}'(d,n)=\lim_{\varepsilon\to 0}\frac{1}{\varepsilon^2}\Big(\sum_{j=1}^{d-2}(|0\rangle+\varepsilon|j\rangle+\frac{\varepsilon^2}{d-2}|d-1\rangle)^{\otimes n}-\frac{1}{\varepsilon}\big(|0\rangle+\varepsilon^2\sum_{j=1}^{d-2}|j\rangle\big)^{\otimes n}-(d-2-\frac{1}{\varepsilon})|0\rangle^{\otimes n}\Big)\,,
\end{equation}
\begin{equation}\label{N-state-BR}
\mathcal{N}'(d,n)=\lim_{\varepsilon\to 0}\frac{1}{\varepsilon}\Big(\sum_{j=1}^{d-1}(|0\rangle+\varepsilon|j\rangle)^{\otimes (n-1)} \otimes |j\rangle + |0\rangle^{\otimes (n-1)}\otimes (\varepsilon|0\rangle - \sum_{j = 1}^{d - 1}|j\rangle)\Big)\,.
\end{equation}

An immediate result of \Cref{theo:LMN-brank} is that the multiqudit $\LL$, $\M$, and $\N$ states are geometrically in the orbit closure of the multiqudit $\GHZ$ states, similarly to how the $\W$ states lie in the orbit closure of the $\GHZ$ states. Again, we see that our families of minimal-rank persistent tensors can be considered as generalizations of multiqubit $\W$ states within multiqudit systems.


\section{Multiqudit Entanglement Transformation}\label{Sec.iv}
Here, we study the SLOCC interconversion between the $n$ -qudit $\GHZ$ state and each generalization of the $\W$ state, that is, the $n$ -qudit $\LL$, $\M$, and $\N$ states. Concerning the chain of degenerations between $\mathcal{L}(d,n)$, $\mathcal{M}(d,n)$, and $\mathcal{N}(d,n)$ in Eq. \eqref{LMN-degenerations}, we are able to study the asymptotic SLOCC transformation between them.

The following proposition relates the Schmidt rank to the asymptotic SLOCC transformation \cite{VC15}.
\begin{proposition}[Ref. \cite{VC15}]\label{prop:rate-SchmidtRank}
Let $|\psi\rangle\in V_1\otimes\cdots\otimes V_n$ be an $n$-partite quantum state and let $\rk_S(\psi)$ denotes the Schmidt rank of $|\psi\rangle$ for any bipartite cut (bipartition) $S|\overline{S}$ where $S\subseteq[n]$ and $\overline{S}=[n]\setminus S$. For any bipartitions, we have
\begin{equation}\label{rate-SchmidtRank}
\omega(\psi,\varphi)\geq\max_{S\subseteq[n]}\frac{\log\rk_S(\varphi)}{\log\rk_S(\psi)}\,.
\end{equation}
\end{proposition}

The following theorem relates degeneration to the asymptotic SLOCC transformation (see Ref. \cite{VC15} for a proof).

\begin{theorem}[Ref. \cite{VC15}]\label{theo:rate-degenration}
Let $|\psi\rangle$ and $|\varphi\rangle$ be two n-partite quantum states. If $|\psi\rangle$ degenerates into $|\varphi\rangle$ via SLOCC, then $\omega(\psi,\varphi)\leq 1$.
\end{theorem}

\begin{theorem}\label{theo:SLOCC-LMN}
An $n$-qudit $\LL$ state can be transformed into an $n$-qudit $\M$ state by asymptotic SLOCC with rate one. An $n$-qudit $\M$ state can be transformed into an $n$-qudit $\N$ state by asymptotic SLOCC with rate one. Formally,
\begin{align}\label{rate-LM}
&\omega(\mathcal{L}(d,n),\mathcal{M}(d,n))=1\,,
\\ \label{rate-MN}
&\omega(\mathcal{M}(d,n),\mathcal{N}(d,n))=1\,.
\end{align}
\end{theorem}
{\it Proof.}
The Schmidt rank of the $n$-qudit $\LL$, $\M$, and $\N$ states across any bipartition is $d$. Therefore, with respect to \Cref{prop:rate-SchmidtRank}, the rates of the aforementioned asymptotic SLOCC transformations are greater than one. From \Cref{theo:LMN-degenerations} we conclude that the upper bounds of the both rates are one. This concludes the proof.
\qed

\begin{corollary}
Based on \Cref{theo:SLOCC-LMN}, we can conclude
\begin{equation}\label{rate-LN}
\omega(\mathcal{L}(d,n),\mathcal{N}(d,n))=1\,.
\end{equation}
\end{corollary}

\begin{theorem}\label{theo:SLOCC-GHZ-L}
An $n$-qudit $\GHZ$ state can be transformed into an $n$-qudit $\LL$ state by asymptotic SLOCC with rate one, i.e.,
\begin{equation}\label{rate-GHZ-L}
\omega(\mathcal{G}(d,n),\mathcal{L}(d,n))=1\,.
\end{equation}
\end{theorem}
{\it Proof.}
The Schmidt rank of the $n$-qudit $\GHZ$ states across any bipartition is $d$. Therefore, with respect to \Cref{prop:rate-SchmidtRank}, the rate of the aforementioned asymptotic SLOCC transformation is greater than one. In the view of \Cref{theo:LMN-degenerations} and \Cref{theo:LMN-brank}, we have the following chain of degenerations
\begin{equation}\label{LMN-degenerations-restate}
\mathcal{G}(d,n)\xdashrightarrow[]{\text{SLOCC}}\mathcal{L}(d,n)\xdashrightarrow[]{\text{SLOCC}}\mathcal{M}(d,n)\xdashrightarrow[]{\text{SLOCC}}\mathcal{N}(d,n)\,.
\end{equation}
Therefore, the upper bound of the rate is one. This concludes the proof.
\qed

\begin{corollary}
Concerning \Cref{theo:SLOCC-LMN} and \Cref{theo:SLOCC-GHZ-L}, one can conclude the following results
\begin{align}\label{rate-GHZ-M}
&\omega(\mathcal{G}(d,n),\mathcal{M}(d,n))=1\,,
\\ \label{rate-GHZ-N}
&\omega(\mathcal{G}(d,n),\mathcal{N}(d,n))=1\,.
\end{align}
\end{corollary}


\section{Direct sums of persistent tensors}
The lower bound obtained in~\Cref{theo:PTLB} can be extended to direct sums with persistent summands and to even more general combinations of tensors, which we call \emph{block pyramidal tensors}. Our lower bound techniques for direct sums and pyramidal tensors generalize some of the constructions of Alder and Strassen in Ref. \cite{AS81} to multipartite tensors. Recent work of Buczy\'{n}ski et al. in Ref. \cite{BPR20} uses similar ideas to prove the rank additivity for some tripartite tensors using the substitution method.

\begin{theorem}\label{theo:PTLB-DirectSum}
Let $\mathcal{T}\in U_1\otimes\cdots\otimes U_n$ be an arbitrary tensor of rank $\rk(\mathcal{T})$. If $\mathcal{P}\in V_1\otimes\cdots\otimes V_n$ is a persistent tensor and $d_k=\dim V_k$, then
\begin{equation}\label{PTLB-DirectSum}
\rk(\mathcal{T}\oplus\mathcal{P})\geq\rk(\mathcal{T})+\sum_{k=1}^{n-1}(d_k-1)+1\,.
\end{equation}
\end{theorem}
\begin{proof}
Consider a tensor rank decomposition
\begin{equation}\label{eq:decomp-DirectSum}
\mathcal{T} \oplus \mathcal{P} =\sum_{p=1}^{r}w_1^{(p)}\otimes\cdots\otimes w_n^{(p)}\,,
\end{equation}
where $w_j^{(p)} = u_j^{(p)} + v_j^{(p)}$ with $u_j^{(p)} \in U_j$ and $v_j^{(p)} \in V_j$.

Let $\pi_{V_j} \colon U_j \oplus V_j \to V_j$ be the canonical projection onto $V_j$ and let $\pi_V=\pi_{V_1}\otimes\cdots \otimes\pi_{V_n}$
Applying $\pi_V$ to the both sides of the decomposition, we get
\begin{equation}
\mathcal{P} =\sum_{p=1}^{r} v_1^{(p)}\otimes\cdots\otimes v_n^{(p)}\,.
\end{equation}

By \Cref{theo:PTLB} we can assume that for $j < n$, $\{v_j^{(D_j+1)},\ldots,v_j^{(D_j+d_j)}\}$ forms a basis of $V_j$, where $D_j = \sum_{k = 1}^{j - 1} (d_k - 1)$.

Let $\pi_{U_n} \colon U_n\oplus V_n \to U_n$ be the canonical projection onto $U_n$. Define the new projections $\Pi_j \colon U_j \oplus V_j \to U_j$ for $j<n$ as
\begin{align}
\begin{cases}
\Pi_j v_j^{(D_j+k)}=-u_j^{(D_j+k)}\,, & \text{$1\leq k\leq d_j$}\,, \\
\Pi_j u=u\,, & \text{$u\in U_j$}\,,
\end{cases}
\end{align}
so that we have $\Pi_j w_j^{(D_j+1)}=\cdots=\Pi_j w_j^{(D_j+d_j)}=0$.
Let $\Pi=\Pi_1\otimes\cdots\otimes\Pi_{n-1}\otimes\pi_{U_n}$.

Note that $\Pi$ sends the first $s=\sum_{k=1}^{n-1}(d_k-1)+1$ summands of the decomposition in Eq.~\eqref{eq:decomp-DirectSum} to zero. Furthermore, $\Pi(\mathcal{T}\oplus\mathcal{P}) =\mathcal{T}$. Therefore, applying $\Pi$ to both sides of the decomposition in Eq.~\eqref{eq:decomp-DirectSum}, one gets a rank decomposition for $\mathcal{T}$ with $r-s$ summands, so $r\geq\rk(\mathcal{T})+s$.
\end{proof}

As a corollary, we find that the rank is additive for direct sums with persistent tensors of minimum rank. Similar rank additivity statements are known for tripartite tensors, the rank of which can be determined using the substitution method~\cite{LM17,BPR20}.

\begin{corollary}
Let $V_1,\ldots,V_n$ be vector spaces with $\dim V_k = d_k$. If $\mathcal{P} \in V_1\otimes\cdots\otimes V_n$ is a persistent tensor of the minimum possible rank $\rk(\mathcal{P}) = \sum_{k=1}^{n-1}(d_k-1)+1$, then for any arbitrary $n$-partite tensor $\mathcal{T}$ we have $\rk(\mathcal{T}\oplus\mathcal{P})=\rk(\mathcal{T})+\rk(\mathcal{P})$.
\end{corollary}

Furthermore, the tensor rank is multiplicative under the Kronecker and tensor products of the persistent tensor of the minimum rank with $\GHZ$ tensors.

\begin{lemma}\label{lem:GHZ-P}
Let $V_1,\ldots,V_n$ be vector spaces with $\dim V_k=d_k$. If $\mathcal{P} \in V_1\otimes\cdots\otimes V_n$ is a persistent tensor of the minimum possible rank $\rk(\mathcal{P})=\sum_{k=1}^{n-1}(d_k-1)+1$, then
\begin{equation}
\rk(\mathcal{G}(d, n)\boxtimes\mathcal{P})=\rk(\mathcal{G}(d, n)\otimes\mathcal{P})=d\cdot\rk(\mathcal{P})\,.
\end{equation}
\end{lemma}
\begin{proof}
Regarding Eq. \eqref{rank-inequalities}, we have $\rk(\mathcal{G}(d, n) \boxtimes \mathcal{P}) \leq \rk(\mathcal{G}(d, n) \otimes \mathcal{P}) \leq d \cdot \rk(\mathcal{P})$.

To get a lower bound, note that the tensors $\mathcal{G}(d, n) \boxtimes \mathcal{P}$ and $\mathcal{P}^{\oplus d}$ are (isometrically) equivalent.
More specifically, $\mathcal{P}^{\oplus d}$ is transformed into $\mathcal{G}(d,n)\boxtimes\mathcal{P}$ by applying to each tensor factor the isomorphism $V_j^{\oplus d} \xrightarrow{\sim} \mathbbm{C}^d \otimes V_j$ sending $(v_0,\ldots,v_{d-1})$ to $\sum_{k=0}^{d-1} e_k \otimes v_k$.

We then apply the previous corollary repeatedly to the direct sum $\mathcal{P}^{\oplus d}$ to obtain the lower bound of the rank $d\cdot\rk(\mathcal{P})$.
\end{proof}

\begin{corollary}\label{cor:Tensor-product-GHZ-W-M-L}
From \Cref{lem:GHZ-P} we have the following corollaries:
\begin{enumerate}
    \item $\rk\big(\mathcal{G}(d,n)\boxtimes\mathcal{W}_n\big)=\rk\big(\mathcal{G}(d,n)\otimes\mathcal{W}_n\big)=nd$.
    \item $\rk\big(\mathcal{G}(d,n)\boxtimes\mathcal{W}_n^{\boxtimes 2}\big)=\rk\big(\mathcal{G}(d,n)\otimes\mathcal{W}_n^{\boxtimes 2}\big) = (3n - 2)d$.
    \item $\rk\big(\mathcal{G}(d_1,n)\boxtimes\mathcal{L}(d_2,n)\big)=\rk\big(\mathcal{G}(d_1,n)\otimes\mathcal{L}(d_2,n)\big)=d_1\big((n-1)d_2-n+2\big)$.
    \item $\rk\big(\mathcal{G}(d_1,n)\boxtimes\mathcal{M}(d_2,n)\big)=\rk\big(\mathcal{G}(d_1,n)\otimes\mathcal{M}(d_2,n)\big)=d_1\big((n-1)d_2-n+2\big)$.
    \item $\rk\big(\mathcal{G}(d_1,n)\boxtimes\mathcal{N}(d_2,n)\big)=\rk\big(\mathcal{G}(d_1,n)\otimes\mathcal{N}(d_2,n)\big)=d_1\big((n-1)d_2-n+2\big)$.
\end{enumerate}
\end{corollary}

In particular, for $\mathcal{G}(d,n) \boxtimes \mathcal{W}_n$ we answer an open question posed in Ref. \cite{CF18}.

The same lower bound method can be applied not only to direct sums but also to a more general class of block tensors.

\begin{definition}
A tensor $\mathcal{Q} \in (U_1 \oplus V_1) \otimes\cdots\otimes(U_n \oplus V_n)$ is a {\bf\emph{block pyramidal tensor}} if $\mathcal{Q}\in U_1\otimes\cdots\otimes U_n\oplus(U_1\oplus V_1)\otimes\cdots\otimes(U_{n-1}\oplus V_{n-1})\otimes V_n$.
Denote by $\pi_{U_k}$ and $\pi_{V_k}$, the canonical projections associated with the summands of $U_k \oplus V_k$.
The tensor $(\pi_{U_1} \otimes \dots \otimes \pi_{U_n}) \mathcal{Q}$ is called the {\bf\emph{head block}} of $\mathcal{Q}$, and $(\pi_{V_1} \otimes \dots \otimes \pi_{V_n}) \mathcal{Q}$ is called the {\bf\emph{step block}} of $\mathcal{Q}$.
\end{definition}

\begin{theorem}\label{theo:PTLB-pyramidal}
Let $\mathcal{Q} \in (U_1 \oplus V_1) \otimes\cdots\otimes(U_n \oplus V_n)$ be a block pyramidal tensor with the head block $\mathcal{T}\in U_1\otimes\cdots\otimes U_n$ and the step block $\mathcal{P}\in V_1\otimes\cdots\otimes V_n$.
If $\mathcal{P}$ is a persistent tensor and $d_k=\dim V_k$, then
\begin{equation}\label{PTLB-pyramidal}
\rk(\mathcal{Q})\geq\rk(\mathcal{T})+\sum_{k=1}^{n-1}(d_k-1)+1\,.
\end{equation}
\end{theorem}
\begin{proof}
The proof is the same as for~\Cref{theo:PTLB-DirectSum}, with $\mathcal{T} \oplus \mathcal{P}$ replaced by $\mathcal{Q}$.
\end{proof}


\section{Conclusion and Outlook}\label{Sec.v}

A central problem in quantum information theory concerns the interconversion between different resources by SLOCC and asymptotic SLOCC. The tensor rank, known as the generalized Schmidt rank, plays an important role in the study of the classification and transformation of multipartite entanglement. Furthermore, the tensor border rank is powerful in studying degeneration and asymptotic SLOCC transformation.

In this work, we have introduced a new class of tensors in $\otimes_{i=1}^n\mathbbm{C}^{d_i}$ that we call persistent tensors and have constructed a lower bound for their tensor rank. We also have introduced several families of persistent tensors in $\otimes^n\mathbbm{C}^d$ where the lower bound is tight. Moreover, we have studied the asymptotic SLOCC transformation of these families of minimal-rank persistent tensors to each other. Showing that the border rank of these families of minimal-rank persistent tensors is $d$, we have concluded that geometrically they are in orbit closure of the $n$-qudit $\GHZ$ states, and we can consider them as generalizations of the $n$-qubit $\W$ states within multiqudit systems. Consequently, we have shown that these generalizations of the $\W$ state can be approximated with arbitrary precision by states in the $\GHZ$ orbit via asymptotic SLOCC. Actually, we have shown that the rate of asymptotic SLOCC transformation from an $n$-qudit $\GHZ$ state into each generalization of the $\W$ state is one. Furthermore, we have proved that the achieved lower bound can be extended to direct sums with persistent summands and to even more general combinations of tensors, which we call block pyramidal tensors. we show that the tensor rank is multiplicative under the Kronecker product and the tensor product of minimal-rank persistent tensors with the GHZ tensor. Although we conjecture that the Kronecker product of persistent tensors is still persistent, we leave the proof as an open problem (see \cref{AppxB}).

Concerning persistent tensors as a new class of tensors with a lower bound of the tensor rank, it would be interesting not only to study this class of tensors more deeply from the complexity theory point of view, but also to study their properties concerning their application in quantum technology. We believe that any application of multiqubit $\W$ states can be studied for its generalizations within multiqudit systems. Indeed, qudit provides several advantages over qubit. For instance, using qudits as building blocks of a quantum circuit can provide a reduction in circuit complexity, since they provide a larger Hilbert space and hence a larger capacity to store and process information \cite{WHSK20}. 
In addition, several benefits of qudits have been proposed, including better noise resistance, higher information coding capacity, stronger nonlocality, enhanced security and more efficient circuit synthesis have been proposed \cite{CBKG02,DWS03,SS11,CMM12,ZZLZCLZ17,FCTECT23}.


\section*{Acknowledgments}
We would like to thank Matthias Christandl for helpful discussions and anonymous reviewers for their comments. Part of the work was done while M.\,G. was at the University of Camerino and V.\,L. was at the University of Copenhagen.
M.\,G. acknowledges the hospitality of the QMATH at the University of Copenhagen where this work was carried out. M.\,G. also acknowledges financial support from the CNR-INO grant agreement n. 4125 and the European Commission through the H2020 QuantERA ERA-NET Co-fund in Quantum Technologies project ``MENTA''.
V.\,L. acknowledges financial support from VILLUM FONDEN via the QMATH Centre of Excellence (Grant No. 10059) and the European Union (ERC Grant Agreements 818761 and 101040907).
Views and opinions expressed are however those of the author(s) only and do not necessarily reflect those of the European Union or the European Research Council Executive Agency. Neither the European Union nor the granting authority can be held responsible for them. 


\bibliographystyle{plain}


\onecolumn\newpage
\appendix


\section{Alternative proof of \Cref{theo:PTLB}}\label{AppxA}

Due to the following lemma, which is the essence of the substitution method (see also Ref. \cite[Appx.~B]{AFT11}) we have the following alternative proof of \Cref{theo:PTLB}.

\begin{lemma}\label{lem:substitution-method}
Let $\mathcal{T}\in V_1\otimes\cdots\otimes V_n$ be an $i$-concise tensor. For every subspace $V'_i\subsetneq V_i$ there exists a projection $\pi_i\colon V_i\to V'_i$ such that 
\begin{equation}\label{substitution-method}
\rk(\mathcal{T})-\rk(\pi_i\mathcal{T})\geq\dim{V_i}-\dim{V'_i}\,,
\end{equation}
where $\pi_i\mathcal{T}$ denotes the application of $\pi_i$ to the $i$-th factor of the tensor $\mathcal{T}$, i.e., $(\mathbb{1}^{\otimes(i-1)}\otimes\pi_i\otimes\mathbb{1}^{\otimes(n-i)})\mathcal{T}$.
\end{lemma}
{\it Proof.}
Suppose Eq. \eqref{rank-decomposition} is a tensor rank decomposition of $\mathcal{T}$, i.e., $\rk(\mathcal{T})=r$. By \Cref{lem:concise-rank} the vectors $\{v_i^{(1)},\ldots,v_i^{(r)}\}$ span $V_i$. Thus, there exists a subset $S_i\subset\{v_i^{(1)},\ldots,v_i^{(r)}\}$ consisting of $c_i=\dim{V_i}-\dim{V'_i}$ vectors such that $W_i=\Span\{S_i\}$ is complementary to $V'_i$. Consider the projection $\pi_i$ onto $V'_i$ along $W_i$. Applying it to the $i$-th factor of each summand of the tensor rank decomposition in Eq. \eqref{rank-decomposition} we obtain a decomposition of $\pi_i\mathcal{T}$ with at most $r - c_i$ summands, because the summands containing vectors from $S_i$ are sent to $0$. It follows that $\rk(\pi_i\mathcal{T})\leq\rk(\mathcal{T})-(\dim{V_i}-\dim{V'_i})$, and we obtain the required statement by rearranging the terms.
\qed

Therefore, the prroof of \Cref{theo:PTLB} can be given as follows:

{\it Proof.}
We prove the statement by induction.

If $n = 2$, then $\mathcal{P}$ is persistent iff it is $1$-concise. By \Cref{cor:concise-matrix} we have $\rk(\mathcal{P}) = d_1$, so the required lower bound is maintained.

Consider now the case $n > 2$. Since $\mathcal{P}$ is a persistent tensor, by \Cref{lem:persistence-equivalent} there exists a vector $|e\rangle\in V_1$ such that for every covector $\langle f|$ in the dual space of $V_1$, $\langle f|\mathcal{P}$ is a persistent tensor whenever $\langle f|e\rangle\neq 0$. Let $V'_1=\Span\{|e\rangle\}$ be a $1$-dimensional subspace of $V_1$. Apply \Cref{lem:substitution-method} to find the projection $\pi_1\colon V_1\to V'_1$ such that $\rk(\mathcal{P})-\rk(\pi_1\mathcal{P})\geq d_1-1$. Since $V'_1$ is a $1$-dimensional subspace, $\pi_1\mathcal{P}=|e\rangle\otimes\mathcal{P}'$ for some $\mathcal{P}'\in V_2\otimes\cdots\otimes V_n$. It follows that $\rk(\mathcal{P}')=\rk(\pi_1\mathcal{P})$ and thus $\rk(\mathcal{P})\geq d_1-1+\rk(\mathcal{P}')$. Note that $\mathcal{P}'=\langle f|\mathcal{P}$ where $\langle f|\in V_1^{\vee}$ is the composition of $\pi_1$ with the linear map $V'_1\to\mathbbm{C}$ that sends $|e\rangle$ to $1$. So, we have $\langle f|e\rangle=1$ and $\mathcal{P}'$ is a persistent tensor.
By the induction hypothesis, we have $\rk(\mathcal{P}')\geq\sum_{k=2}^{n-1}(d_k-1)+1$, and therefore $\rk(\mathcal{P})\geq d_1-1+\rk(\mathcal{P}')\geq\sum_{k=1}^{n-1}(d_k-1)+1$.
\qed


\section{Is the Kronecker product of persistent tensors still persistent?}\label{AppxB}

Although we have checked many examples that the Kronecker product of two persistent tensors is a persistent tensor, we leave the proof as an open problem.

\begin{conjecture}\label{conj:P1P2}
Let $U_1,\ldots,U_n$ and $V_1,\ldots,V_n$ be vector spaces with $\dim U_k=d_k$ and $\dim V_k=d'_k$, respectively. If $\mathcal{P}_1\in U_1\otimes\cdots\otimes U_n$ and $\mathcal{P}_2\in V_1\otimes\cdots\otimes V_n$ are two persistent tensors, then their Kronecker product is also a persistent tensor and therefore,
\begin{equation}
\rk(\mathcal{P}_1\boxtimes\mathcal{P}_2)\geq\sum_{k=1}^{n}(d_k+d'_k-1)+1\,.
\end{equation}
\end{conjecture}


\end{document}